\documentclass{bcp}

\def\LaTeX{L\kern -.36em\raise .3ex\hbox{\sc a}\kern -.15em T\kern -.1667em%
\lower .7ex\hbox{E}\kern -.125em X}

\begin{document}

%%\keywords{These are optional}
\mathclass{ }
%%\thanks{Research of the first author supported by KBN grant 00-000.}
\abbrevauthors{P.-H. Chavanis}
\abbrevtitle{Generalized kinetic equations}

\title{Generalized kinetic equations and\\
effective thermodynamics}

\author{Pierre-Henri Chavanis}
\address{Laboratoire de Physique Th\'eorique, Universit\'e
Paul Sabatier\\ 118 route de Narbonne, 31062 Toulouse Cedex 4,
France\\
E-mail: chavanis@irsamc.ups-tlse.fr}

\maketitlebcp

\abstract{We introduce a new class of nonlocal kinetic equations 
and nonlocal Fokker-Planck equations associated with an effective
generalized thermodynamical formalism. These equations have a rich
physical and mathematical structure that can describe phase
transitions and blow-up phenomena. On general grounds, our formalism
can have applications in different domains of physics, astrophysics,
hydrodynamics and biology. We find an aesthetic connexion between
topics (stars, vortices, bacteries,...)  which were previously
disconnected.  The common point between these systems is the
(attractive) long-range nature of the interactions.}

\section*{1. Introduction.} 

The statistical mechanics of systems with long-range interactions is
currently a topic of active research in physics
\cite{dauxois}. Systems with long-range interactions are numerous in
nature: self-gravitating systems, two-dimensional vortices,
non-neutral plasmas, metallic clusters, dipoles, fracture etc. These
systems exhibit similar features such as negative specific heats,
inequivalence of statistical ensembles, phase transitions,
self-organization and persistence of metastable states.  Among all the
previous examples, self-gravitating systems and 2D vortices play a
special role because they both interact via an unshielded Newtonian
potential (in dimensions $D=3$ or $D=2$) and possess a rather similar
mathematical structure \cite{houches}. Coincidentally, the chemotactic
aggregation of bacteries in biology has some connexions with the
collapse and organization of self-gravitating systems and 2D vortices
\cite{gt,bcp2}.

For systems with long-range interactions, the mean-field approximation
turns out to be exact in a proper thermodynamic limit
\cite{bbgky}. Therefore, their equilibrium description amounts to
solving a variational problem, namely the maximization of the
Boltzmann entropy $S_{B}[f]$ at fixed mass $M=M[f]$ and energy
$E=E[f]$. In the case of quantum particles, the Boltzmann entropy is
replaced by the Fermi-Dirac entropy $S_{FD}[f]$.  This maximization
problem determines the most probable distribution of particles at
statistical equilibrium, assuming that all the accessible microstates
(with given $E$ and $M$) are equiprobable. However, if, for some
reason, the microstates are {\it not} equiprobable, other forms of
entropy can emerge. This appears in the physics of complex media
displaying a fractal (or multifractal) structure and exhibiting
anomalous diffusion. In general, these effects are due to the
existence of ``hidden constraints'' \cite{gtkin} that change the form
of the transitions probabilities that we would naively expect. 
Motivated by this problem, we shall develop an {\it effective}
generalized thermodynamical formalism (in $\mu$ space) associated with
a larger class of entropy functionals. Specifically, we consider the
maximization of a generalized entropy $S[f]=-\int C(f) d^{D}{\bf
r}d^{D}{\bf v}$, where $C$ is a convex function, at fixed mass $M$ and
energy $E$. Since the energy is fixed, we call this a microcanonical
description. We also consider the minimization of a generalized free
energy $F=E-TS$ at fixed mass $M$ and temperature $T$. Since the
temperature is fixed, we call this a canonical description. We discuss
the inequivalence of these two descriptions when the caloric curve
$\beta (E)$ presents turning points or bifurcations. This occurs in
particular when the potential of interaction is long-ranged.

We also introduce a new class of relaxation equations associated with
this generalized thermodynamical formalism \cite{gt,gtkin,lemou}. We
first consider a generalized class of Fokker-Planck equations
extending the ordinary Kramers and Smoluchowski equations
\cite{gt}. These equations have a canonical structure as they decrease
a generalized free energy.  When the potential of interaction is
long-ranged, these equations are non-local in space and can exhibit a
rich variety of behaviors including phase transitions and blow-up
phenomena. In the limit of short-ranged interactions, they reduce to
Cahn-Hilliard equations \cite{lemou}.  We also consider a generalized
class of kinetic equations extending the ordinary Boltzmann and Landau
equations \cite{gtkin}. These kinetic equations have a microcanonical
structure as they conserve energy and increase a generalized entropy.

Our formalism can have applications in different domains of physics,
astrophysics, fluid mechanics, biology, economy etc. with various
interpretations that are not necessarily connected to
thermodynamics. It is therefore important to develop a general
formalism without reference to a specific context. Then, a
justification has to be given in each case. For example, the
maximization of the functional $S[f]=-\int C(f)d^{D}{\bf r}d^{D}{\bf
v}$, where $C$ is convex, at fixed mass and energy determines a
nonlinearly dynamically stable stationary solution of the
Vlasov-Poisson system for collisionless stellar systems (and 2D
vortices). In this context, $S[f]$ is a H-function
\cite{tremaine}, not an entropy. Then, our generalized relaxation
equations \cite{gt} can be used as numerical algorithms to construct
stable stationary solutions of the Vlasov-Poisson system. On the other
hand, equations similar to generalized Fokker-Planck equations 
appear in biology in relation with the chemotactic aggregation of
bacterial populations. In any case, it is useful to develop a {\it
thermodynamical analogy} \cite{gt} and use a vocabulary borrowed from
thermodynamics. Thus, we can directly transpose the methods developed
in thermodynamics to a different context.

\section*{2. Maximum entropy principle} Let us consider a system of $N$
particles in interaction and denote by $f({\bf r},{\bf v},t)$
their distribution function defined such that $f d^{D}{\bf
r}d^{D}{\bf v}$ gives the mass of particles with position
${\bf r}$ and velocity ${\bf v}$ at time $t$. The spatial density
is
\begin{equation}
\label{mep1} \rho({\bf r},t)=\int f({\bf r},{\bf v},t) \ d^{D}{\bf
v},
\end{equation}
and the total mass
\begin{equation}
\label{mep2} M=\int \rho({\bf r},t) \ d^{D}{\bf r}.
\end{equation}

Let ${\bf F}({\bf
r},t)=-\nabla\Phi$ be the force (by unit of mass) experienced by a
particle. We assume that the potential $\Phi({\bf r},t)$ is
related to the density $\rho({\bf r},t)$ by a relation of the form
\begin{equation}
\label{mep3}\Phi({\bf r},t)=\int \rho({\bf r}',t)u({\bf r}-{\bf
r'})d^{D}{\bf r}',
\end{equation}
where $u({\bf r}-{\bf r'})$ is an arbitrary binary potential
depending only on the absolute distance $|{\bf r}-{\bf r'}|$
between the particles. The energy can be expressed as
\begin{equation}
\label{mep4} E=\int{1\over 2}f v^{2} d^{D}{\bf  r}
 d^{D}{\bf  v}+{1\over 2}\int \rho \Phi \ d^{D}{\bf  r}=K+W,
\end{equation}
where $K$ is the kinetic energy and $W$ the potential energy. The
following results remain valid if $\Phi=\Phi_{ext}({\bf r})$ is a
fixed external potential, in which case the potential energy reads $W=\int
\rho\Phi_{ext}d^{D}{\bf r}$.

We introduce a generalized entropy of the form
\begin{equation}
\label{mep5} S=-\int C(f)\ d^{D}{\bf r}d^{D}{\bf v},
\end{equation}
where $C(f)$ is a convex function, i.e. $C''(f)> 0$. We are
interested by the distribution function $f({\bf r},{\bf v})$ which
maximizes the generalized entropy (\ref{mep5}) at fixed mass and
energy, i.e.
\begin{equation}
\label{mep6} {\rm Max}\quad S[f]\quad | \quad E[f]=E,\ M[f]=M.
\end{equation}
Since the energy is fixed, we shall associate this maximization
problem to a microcanonical description. Introducing Lagrange
multipliers and writing the variational principle in the form
\begin{equation}
\label{mep7}
\delta S-\beta\delta E-\alpha\delta M=0,
\end{equation}
we find that the critical points of entropy at fixed mass and energy
are given by
\begin{equation}
\label{mep8}
C'(f)=-\beta\epsilon-\alpha,
\end{equation}
where $\epsilon={v^{2}\over 2}+\Phi({\bf r})$ is the energy of a
particle by unit of mass.  The Lagrange multipliers $\beta=1/T$
and $\alpha$ can be interpreted as a generalized inverse
temperature and a generalized chemical potential, respectively.
Equation (\ref{mep8}) can be written equivalently as
\begin{equation}
\label{mep9}
f=F(\beta\epsilon+\alpha),
\end{equation}
where $F(x)=(C')^{-1}(-x)$. From the identity
\begin{equation}
\label{mep10}
f'(\epsilon)=-\beta/C''(f),
\end{equation}
resulting from Eq. (\ref{mep8}), we find that $f(\epsilon)$ is a
monotonically decreasing function of energy if $\beta>0$. The case of
negative temperatures $\beta<0$ can also be of interest depending on
the form of the function $C$. Explicating the relation between the
potential and the density, the equilibrium distribution is determined
by the integro-differential equation
\begin{equation}
\label{mep11} C'(f)=-\beta\biggl\lbrace {v^{2}\over 2}+\int f({\bf r}',{\bf v}')u({\bf r}-{\bf r}')d^{D}{\bf r}'d^{D}{\bf v}'\biggr\rbrace-\alpha.
\end{equation}
The conservation of angular momentum ${\bf L}=\int f {\bf r}\times
{\bf v} d^{3}{\bf r}d^{3}{\bf v}$ (in $D=3$) can be easily included in
the variational principle (\ref{mep7}) by introducing an appropriate
Lagrange multiplier $\beta {\bf \Omega}$, where ${\bf \Omega}$ is the
angular velocity. Equation (\ref{mep8}) remains valid provided that
$\epsilon$ is replaced by the Jacobi energy
$\epsilon_{J}=\epsilon-{\bf\Omega}\cdot ({\bf r}{\times}{\bf
v})={1\over 2}{\bf w}^{2}+\Phi_{eff}$ where ${\bf w}={\bf
v}-{\bf\Omega}{\times}{\bf r}$ is the relative velocity and
$\Phi_{eff}=\Phi-{1\over 2}({\bf\Omega}{\times} {\bf r})^{2}$ is the
{\it effective} potential accounting for inertial forces.

We now introduce the generalized free energy
\begin{equation}
\label{mep12} J[f]=S[f]-\beta E[f],
\end{equation}
associated with the functionals (\ref{mep4}) and (\ref{mep5}). We are
interested by the distribution function $f({\bf r},{\bf v})$ which
maximizes the generalized free energy (\ref{mep12}) at fixed mass and
temperature, i.e.
\begin{equation}
\label{mep13} {\rm Max}\quad J[f]=S[f]-\beta E[f]\quad | \quad
M[f]=M.
\end{equation}
Since the temperature is given, we shall associate this
maximization problem to a canonical description. Introducing
Lagrange multipliers and writing the variational principle in the
form
\begin{equation}
\label{mep14} \delta J-\alpha\delta M=0,
\end{equation}
we find that the critical points of free energy at fixed mass and
temperature are given by
\begin{equation}
\label{mep15} C'(f)=-\beta\epsilon-\alpha,
\end{equation}
as in the microcanonical description.

\section*{3. Stability conditions} The critical points of the variational 
problems (\ref{mep6}) and (\ref{mep13}) are the same. The equilibrium
state is then obtained by solving the integro-differential equation
(\ref{mep11}) and relating the temperature $\beta$ to the energy
$E$. We can thus plot the generalized caloric curve $\beta(E)$
parameterizing the series of equilibria. We need now to determine the
stability of these solutions by investigating the sign of the second
order variations of $S$ or $J$. In the microcanonical situation, we
must select {\it maxima} of $S[f]$ at fixed mass and energy. The
condition that $f$ is a {maximum} of $S$ at fixed mass and energy is
equivalent to the condition that $\delta^{2}{J}\equiv
\delta^{2}{S}-\beta\delta^{2}E$ is negative for all perturbations
that conserve mass and energy to first order. This condition can
be written
\begin{eqnarray}
\label{ts1}
\delta^{2}J=-\int C''(f){(\delta f)^{2}\over 2}d^{D}{\bf  r} d^{D}{\bf  v}-{1\over 2}\beta\int \delta\rho\delta\Phi d^{D}{\bf  r}\le 0,\nonumber\\
\forall\ \delta f \mid\ \delta E=\delta M=0.\qquad\qquad \qquad
\end{eqnarray}
The condition of stability in the canonical situation requires
that $f$ is a {\it maximum} of $J[f]$ at fixed mass and
temperature. This is equivalent to the condition that
$\delta^{2}{J}$ is negative for all perturbations that conserve
mass. This can be written
\begin{eqnarray}
\label{ts2}
\delta^{2}J=-\int C''(f){(\delta f)^{2}\over 2}d^{D}{\bf  r} d^{D}{\bf  v}-{1\over 2}\beta\int \delta\rho\delta\Phi d^{D}{\bf  r}\le 0,\nonumber\\
\forall\ \delta f \mid\ \delta M=0.\qquad\qquad \qquad
\end{eqnarray}
Using Eq. (\ref{mep10}), the functional arising in these stability criteria can be expressed as
\begin{eqnarray}
\label{ts3}
\delta^{2}J={1\over 2}\beta\biggl\lbrace \int {(\delta f)^{2}\over f'(\epsilon)}d^{D}{\bf  r} d^{D}{\bf  v}-\int \delta\rho\delta\Phi d^{D}{\bf  r}\biggr\rbrace.
\end{eqnarray}
We emphasize the importance of the first order constraints in the
stability analysis. We note that canonical stability implies
microcanonical stability but that the converse is wrong in
general. Indeed, if inequality (\ref{ts2}) is satisfied for all
perturbations that conserve mass, it is a fortiori satisfied for
perturbations that conserve mass {\it and} energy. Since the converse
is wrong, this implies that some solutions can be stable in the
microcanonical ensemble while they are unstable in the canonical
one. The microcanonical and canonical descriptions are inequivalent
when the caloric curve $\beta(E)$ presents turning points. This
situation is well-known in the case of systems with long-range
interactions such as self-gravitating systems \cite{pt}. The stability
of the solutions can be decided by using the turning point criterion
of Katz \cite{katz} which extends the theory of Poincar\'e on linear
series of equilibria. It is found that a change of stability in the
series of equilibria occurs in the microcanonical ensemble when the
energy is extremum and in the canonical ensemble when the temperature
is extremum. Stability is lost or gained depending on whether the
series of equilibria $\beta(E)$ turns clockwise or anti-clockwise at
that critical point. An illustration of these results is proposed in
Ref. \cite{pt}, in the case of self-gravitating fermions. A change of
stability along a series of equilibria can also occur at a branching
point \cite{katz}, where the solutions bifurcate. A general
classification of phase transitions for systems with long-range
interactions has been proposed by Bouchet \& Barr\'e
\cite{bb}.

\section*{4. The free energy functional} The maximization problem
(\ref{mep13}) in the canonical ensemble can be simplified. First of all, we
write the free energy in the usual form
\begin{equation}
\label{fef1} F[f]=E[f]-TS[f].
\end{equation}
To solve the minimization problem
\begin{equation}
\label{fef2} {\rm Min}\quad F[f]=E[f]-TS[f]\quad | \quad M[f]=M,
\end{equation}
we shall proceed in two steps. We first minimize $F[f]$ at fixed
density $\rho({\bf r})$. Introducing a Lagrange multiplier
$\lambda({\bf r})$, we find that the global minimum $f_{*}({\bf
r},{\bf v})$ of this variational problem is determined by
\begin{equation}
\label{fef3} C'(f_{*})=-\beta\biggl\lbrack {v^{2}\over 2}+\lambda({\bf r})\biggr\rbrack.
\end{equation}
The distribution function $f_{*}$ can be written
\begin{equation}
\label{fef5} f_{*}=F\biggl \lbrack \beta\biggl ({v^{2}\over 2}+\lambda({\bf r})\biggr )\biggr\rbrack,
\end{equation}
where $F(x)=(C')^{-1}(-x)$. We define the density and the pressure by
\begin{equation}
\label{fef4}\rho=\int f d^{D}{\bf v},\qquad p={1\over D}\int f v^{2}d^{D}{\bf v}.
\end{equation}
Substituting Eq. (\ref{fef5}) in the foregoing expressions,
we find that
\begin{equation}
\label{fef6}\rho={1\over \beta^{D/2}}g(\beta\lambda), \qquad p={1\over \beta^{D+2\over 2}}h(\beta\lambda),
\end{equation}
with
\begin{equation}
\label{fef7} g(x)=2^{D-2\over 2}S_{D} \int_{0}^{+\infty} F(x+t)\ t^{D-2\over 2} dt,
\end{equation}
\begin{equation}
\label{fef8} h(x)={1\over D}2^{D\over 2}S_{D} \int_{0}^{+\infty} F(x+t)\ t^{D\over 2} dt,
\end{equation}
where $S_{D}$ is the surface of a unit sphere in $D$
dimensions. Eliminating $\lambda$ between the foregoing expressions, we find
that the fluid is {\it barotropic}, in the sense that $p=p(\rho)$
where the equation of state is entirely specified by $C(f)$. We can
now express the free energy (\ref{fef1}) as a functional of $\rho$ by using
$F[\rho]=F[f_{*}]$. The energy (\ref{mep4}) is simply given by
\begin{equation}
\label{fef9}
E={D\over 2}\int p\ d^{D}{\bf r}+{1\over 2}\int\rho\Phi d^{D}{\bf
  r}.
\end{equation}
On the other hand, the entropy (\ref{mep5}) can be written
\begin{eqnarray}
{S}=-{2^{D-2\over 2}S_{D}\over\beta^{D/2}}\int d^{D}{\bf
  r}\int_{0}^{+\infty}C\lbrack F(t+\beta\lambda)\rbrack \ t^{D-2\over 2}dt.
\label{fef10}
\end{eqnarray}
Integrating by parts  and using $C'\lbrack F(x)\rbrack=-x$, we find that
\begin{eqnarray}
{S}=-{2^{D/2}S_{D}\over D\beta^{D/2}}\int d^{D}{\bf r}\int_{0}^{+\infty}
F'(t+\beta\lambda)(t+\beta\lambda)t^{D/2}dt.
\label{fef11}
\end{eqnarray}
Integrating by parts one more time and using Eqs. (\ref{fef6}),
(\ref{fef7}) and (\ref{fef8}), we finally obtain
\begin{eqnarray}
{S}={D+2\over 2}\beta\int p d^{D}{\bf r}+\beta\int \lambda\rho d^{D}{\bf r}.
\label{fef12}
\end{eqnarray}
Collecting all the previous results, the free energy (\ref{fef1}) becomes
\begin{eqnarray}
{F}[\rho]=-\int\rho \biggl (\lambda+{p\over\rho}\biggr )d^{D}{\bf
r} +{1\over 2}\int\rho\Phi d^{D}{\bf r}. \label{fef13}
\end{eqnarray}
Finally, using the relation $h'(x)=-g(x)$ obtained from
Eqs. (\ref{fef7}) and (\ref{fef8}) by a simple integration by parts, it
is easy to check that Eq. (\ref{fef6}) implies
\begin{eqnarray}
\lambda+{p\over\rho}=-\int_{0}^{\rho}{p(\rho')\over\rho'^{2}}d\rho'.
\label{fef14}
\end{eqnarray}
Hence, the free energy can be written more explicitly as
\begin{eqnarray}
{F}[\rho]=\int \rho \int_{0}^{\rho}{p(\rho')\over\rho'^{2}}d\rho'
d^D{\bf r}+{1\over
  2}\int\rho\Phi d^{D}{\bf r}.
\label{fef15}
\end{eqnarray}
We are led therefore to study the minimization problem
\begin{equation}
\label{fef16} {\rm Min}\quad F[\rho]\quad | \quad M[\rho]=M,
\end{equation}
for the free energy functional (\ref{fef15}).  Writing the first order
variations in the form
\begin{equation}
\label{fef17} \delta F-\alpha\delta M=0,
\end{equation}
we find that the critical points of free energy satisfy the condition of hydrostatic balance
\begin{equation}
\label{fef18} \nabla p=-\rho\nabla\Phi,
\end{equation}
and that $\lambda=\Phi+\alpha/\beta$. We can easily check that the condition of
hydrostatic equilibrium is directly implied by the relation
$f=f(\epsilon)$ derived in Sec. 2. This immediately results from the
identities
\begin{equation}
\label{fef19} \rho={1\over D}\int f {\partial {\bf v}\over\partial {\bf v}}d^{D}{\bf v}=-{1\over D}\int {\partial f\over\partial {\bf v}}\cdot {\bf v}d^{D}{\bf v}=-{1\over D}\int f'(\epsilon)\ v^{2}d^{D}{\bf v},
\end{equation}
\begin{equation}
\label{fef20} \nabla p={1\over D}\int  {\partial f\over\partial {\bf r}}v^{2}d^{D}{\bf v}={1\over D}\int f'(\epsilon)\nabla\Phi v^{2} d^{D}{\bf v}=-\rho\nabla\Phi.
\end{equation}
Explicating the relation between the potential and the density,
the condition of hydrostatic equilibrium can be written in the
form of an integro-differential equation
\begin{equation}
\label{fef21} {p'(\rho)\over \rho}\nabla \rho=-\nabla\int \rho({\bf r}')u({\bf r}-{\bf r}')d^{D}{\bf r}'.
\end{equation}
This equation determines the equilibrium solutions (critical points of $F$). Their stability (minima of $F$) is determined by the condition
\begin{eqnarray}
\label{fef22}
\delta^{2}F=\int {p'(\rho)\over 2\rho}(\delta \rho)^{2}d^{D}{\bf  r}+{1\over 2}\int \delta\rho\delta\Phi d^{D}{\bf  r}\ge 0,\nonumber\\
\forall\ \delta\rho \mid \delta M=0.\qquad\qquad \qquad
\end{eqnarray}
Using the condition of hydrostatic equilibrium, the above functional can be rewritten as
\begin{eqnarray}
\label{fef23}
\delta^{2}F={1\over 2}\int {(\delta \rho)^{2}\over -\rho'(\Phi)} \ d^{D}{\bf  r}+{1\over 2}\int \delta\rho\delta\Phi \ d^{D}{\bf  r}.
\end{eqnarray}

\section*{5. Examples of entropy functionals}
Among all functionals of the form (\ref{mep5}), some have been discussed in detail in the literature. The most famous functional is the Boltzmann entropy
\begin{equation}
\label{ex1}
S_{B}[f]=-\int f\ln f d^{D}{\bf r}d^{D}{\bf v}.
\end{equation}
It leads to the isothermal distribution
\begin{equation}
\label{ex2}
f=A e^{-\beta\epsilon}.
\end{equation}
The corresponding distribution is physical space is the Boltzmann distribution
\begin{equation}
\label{ex3}
\rho=A' e^{-\beta\Phi},\qquad A'=\biggl ({2\pi\over \beta}\biggr )^{D/2} A.
\end{equation}
The  distribution function (\ref{ex2}) leads to the classical equation of state
\begin{equation}
\label{ex4}
p=\rho T,
\end{equation}
and to the free energy
\begin{eqnarray}
{F}[\rho]=T\int \rho \ln\rho\ d^D{\bf r}+{1\over
  2}\int\rho\Phi \ d^{D}{\bf r}.
\label{ex5}
\end{eqnarray}

Closely related to the Boltzmann entropy is the Fermi-Dirac entropy
\begin{equation}
\label{ex6}
S_{FD}[f]=-\int \biggl\lbrace {f\over \eta_{0}}\ln {f\over \eta_{0}}+\biggl (1-{f\over \eta_{0}}\biggr )\ln \biggl (1-{f\over \eta_{0}}\biggr ) \biggr\rbrace d^{D}{\bf r} d^{D}{\bf v},
\end{equation}
which leads to the Fermi-Dirac distribution function
\begin{equation}
\label{ex7}
f={\eta_{0}\over 1+\lambda e^{\beta\eta_{0}\epsilon}}.
\end{equation}
The Fermi-Dirac distribution function (\ref{ex7}) satisfies the
constraint $f\le \eta_{0}$ which is related to Pauli's exclusion
principle in quantum mechanics. The isothermal distribution function
(\ref{ex2}) is recovered in the non-degenerate limit $f\ll
\eta_{0}$.   The distribution in physical space associated with the Fermi-Dirac statistics is
\begin{equation}
\label{ex8}
\rho={\eta_{0}S_{D}2^{{D\over 2}-1}\over\beta^{D/2}}I_{{D\over 2}-1}(\lambda e^{\beta\Phi}),
\end{equation}
where $I_{n}$ is the Fermi integral
\begin{equation}
\label{ex9}
I_{n}(t)=\int_{0}^{+\infty}{x^{n}\over 1+te^{x}}dx.
\end{equation}
The  distribution function (\ref{ex7}) leads to the quantum equation of state given in parametric form as
\begin{equation}
\label{ex10}
\rho={\eta_{0}S_{D}2^{{D\over 2}-1}\over\beta^{D/2}}I_{{D\over 2}-1}(\lambda'),\qquad p={\eta_{0}S_{D}2^{{D\over 2}}\over D\beta^{{D\over 2}+1}}I_{{D\over 2}}(\lambda').
\end{equation}

Recently, there was a considerable interest in physics for
functionals of the form
\begin{equation}
\label{ex11}
S_{q}[f]=-{1\over q-1}\int (f^{q}-f)  d^{D}{\bf r} d^{D}{\bf v},
\end{equation}
where $q$ is a real number. Such functionals introduced by Tsallis
\cite{tsallis} are called $q$-entropies. They lead to polytropic
distribution functions 
\begin{equation}
\label{ex12}
f=\biggl\lbrack \mu-{(q-1)\beta\over q}\epsilon\biggr\rbrack^{1\over q-1}.
\end{equation}
The index $n$ of the polytrope is related to the
parameter $q$ by the relation $n=D/2+1/(q-1)$ \cite{anomalous}.
Isothermal distribution functions are recovered in the limit
$q\rightarrow 1$ (i.e. $n\rightarrow +\infty$). The distribution
function (\ref{ex12}) leads to the polytropic equation of state
\begin{equation}
\label{ex13}
p=K\rho^{\gamma}, \qquad \gamma=1+{1\over n},
\end{equation}
with
\begin{equation}
\label{ex14}
K={1\over n+1}\biggl\lbrace 2^{{D\over 2}-1}S_{D}AB\biggl ({D\over 2},n+1-{D\over 2}\biggr )\biggr\rbrace^{-1/n},
\end{equation}
where $A=\lbrack (q-1)\beta/q\rbrack^{1\over q-1}$ and $B(n,m)$ is the Beta function. The corresponding free energy can be written 
\begin{eqnarray}
{F}[\rho]={K\over\gamma -1}\int (\rho^{\gamma}-\rho) \ d^D{\bf r}+{1\over
  2}\int\rho\Phi \ d^{D}{\bf r},
\label{ex15}
\end{eqnarray}
and the density in physical space is
\begin{equation}
\label{ex16}
\rho=\biggl\lbrack \lambda-{\gamma-1\over K\gamma}\Phi\biggr\rbrack^{1\over\gamma-1}.
\end{equation}
We note that a polytropic distribution with index $q$ in phase space
yields a polytropic distribution with index $\gamma=1+2(q-1)/\lbrack
2+D(q-1)\rbrack$ in physical space. In this sense, Tsallis
distributions are stable laws. By comparing Eqs. (\ref{ex12}) and
(\ref{ex16}) or Eqs. (\ref{fef1}), (\ref{ex11}) and (\ref{ex15}) we
note that $K$ plays the same role in physical space as the temperature
$T=1/\beta$ in phase space. It is sometimes called a ``polytropic
temperature''.

We have started a systematic study of the variational problems
(\ref{mep6}), (\ref{mep13}) and (\ref{fef16}) by considering first the
gravitational interaction in $D$ dimensions. The case of isothermal
self-gravitating systems associated with the Boltzmann entropy has
been considered in \cite{sire}. The case of self-gravitating fermions
associated with the Fermi-Dirac entropy has been considered in
\cite{pt,fermiD}. The case of polytropic self-gravitating systems
associated with the Tsallis entropy has been considered in
\cite{anomalous}.

\section*{6. Physical applications} The maximization problems discussed previously can have various physical applications (not necessarily related to thermodynamics) that we briefly mention.

{\it (i) Statistical mechanics:} The variational principles
(\ref{mep6}), (\ref{mep13}) and (\ref{fef16}) determine the
statistical equilibrium states of systems with long-range
interactions, such as self-gravitating systems. In that case, $S[f]$
is the Boltzmann entropy (\ref{ex1}) for classical particles (e.g.,
stars in globular clusters) \cite{lbw} or the Fermi-Dirac entropy
(\ref{ex6}) for fermions (e.g., massive neutrinos in dark matter
models) \cite{pt}. The microcanonical situation (\ref{mep6}) applies
to isolated Hamiltonian systems such as stellar systems
\cite{paddy}. The canonical situation (\ref{mep13}) or (\ref{fef16})
applies to systems in contact with a heat bath imposing its
temperature, like for the interstellar medium \cite{aa}. This is also
the proper statistical description of a gas of self-gravitating
Brownian particles \cite{crs,revbd}. This discussion remains valid for other
types of long-range interactions. 

{\it (ii) Vlasov equation:} The variational principle (\ref{mep6})
determines nonlinearly dynamically stable stationary solutions of the
Vlasov-Poisson system
\begin{equation}
\label{ge10}
{\partial f\over\partial t}+{\bf v}\cdot {\partial f\over\partial {\bf r}}+{\bf F}\cdot {\partial f\over\partial {\bf v}}=0,
\end{equation}
\begin{equation}
\label{ge10b}
\Delta\Phi=4\pi G\int f d^{3}{\bf v},
\end{equation}
which describes collisionless stellar systems. These robust organized
states can emerge as a result of a violent relaxation \cite{vr,csr}. In
that context, $S[f]$ is called a H-function \cite{tremaine}. Boltzmann
and Tsallis functionals are particular H-functions (not true entropies
in that context) associated with isothermal and polytropic stellar
systems \cite{grand,gt}. Note that the criterion of nonlinear
dynamical stability (\ref{mep6})-(\ref{ts1}) is more refined than the stability
criterion (\ref{mep13})-(\ref{ts2}). These results on nonlinear dynamical stability
remain valid when the force ${\bf F}$ is related to the
distribution function $f$ by a relation of the general form
(\ref{mep3}).

{\it (iii) Euler-Jeans equation:} The  variational principle (\ref{fef16}) determines the nonlinear dynamical stability of stationary solutions of the Euler-Jeans-Poisson system 
\begin{equation}
{\partial\rho\over\partial t}+\nabla \cdot (\rho {\bf u})=0,
\label{bs1}
\end{equation}
\begin{equation}
{\partial {\bf u}\over\partial t}+({\bf u}\cdot \nabla) {\bf u}=-{1\over\rho}\nabla p-\nabla\Phi, \qquad \Delta\Phi=4\pi G\rho,
\label{bs2}
\end{equation}
describing barotropic stars. In that context, ${\cal
W}[\rho]=F[\rho]+\int \rho {u^{2}\over 2}d^{3}{\bf r}$ represents the
energy of the star. The first term in Eq. (\ref{fef15}) is the
internal energy of a barotropic gas and the second term is the
gravitational energy (a kinetic term has also to be introduced in the
energy functional ${\cal W}$). Since the stability condition
(\ref{fef16}) implies (\ref{mep6}), we note that a stellar system is
stable with respect to the Vlasov equation whenever the corresponding
barotropic star with the same equation of state is stable with respect
to the Euler-Jeans equations (but the reciprocal is wrong in general)
\cite{grand}.  In astrophysics, this is known as the Antonov's first
law \cite{bt}.

{\it (iv) Generalized thermodynamics:} The variational principles
(\ref{mep6}), (\ref{mep13}) and (\ref{fef16}) determine the
generalized thermodynamical stability of complex systems exhibiting
anomalous diffusion. Such systems have a complicated phase space structure
(fractal, multifractal,...) due to the action of microscopic
constraints (hidden constraints) that are often difficult to
formalize. These systems can be described by {\it effective} equations
resembling generalized Boltzmann and Fokker-Planck equations
\cite{gt,gtkin}.  The variational principles (\ref{mep6}),
(\ref{mep13}) and (\ref{fef16}) also determine the linear dynamical
stability of stationary solutions of these equations. Accordingly, there is
a close connexion between thermodynamical and dynamical stability
\cite{gt}.

\section*{7. Generalized Kramers equation} We shall introduce formally a relaxation equation associated with the minimization problem (\ref{fef2}). We write this equation in the form
\begin{equation}
\label{gk1}
{\partial f\over\partial t}+{\bf v}\cdot {\partial f\over\partial {\bf
r}}+{\bf F}\cdot {\partial f\over\partial {\bf v}}={\partial
\over\partial {\bf v}}\cdot \biggl\lbrack D\beta f {\partial
\over\partial {\bf v}} \biggl ({\delta F\over\delta
f}\biggr )\biggr\rbrack,
\end{equation}
where $\delta/\delta f$ denotes a functional derivative. By
construction, this equation conserves mass since the right hand side
is the divergence of a current $-{\bf J}_{f}$. Using the expression of
the free energy (\ref{fef1}) with Eqs. (\ref{mep4}) and (\ref{mep5}),
we obtain the generalized Fokker-Planck equation
\begin{equation}
\label{gk2}
{\partial f\over\partial t}+L f={\partial \over \partial {\bf v}}\cdot \biggl\lbrace D\biggl\lbrack f C''(f){\partial f\over\partial {\bf v}}+\beta f{\bf v}\biggr \rbrack \biggr\rbrace,
\end{equation}
where $L$ is the advection operator. Morphologically, Eq.  (\ref{gk2})
can be viewed as a generalized non-local Kramers equation
\cite{risken}. It describes the dynamics (in phase space) of a
system of Langevin particles in interaction governed by a generalized
class of stochastic processes \cite{gt,bbgky}:
\begin{eqnarray}
\label{a16}
{d{\bf r}_{i}\over dt}={\bf v}_{i},\quad
{d{\bf v}_{i}\over dt}=-\xi{\bf v}_{i}-\nabla_{i}U({\bf r}_{1},...,{\bf r}_{N})+\sqrt{2Df_{i}\biggl\lbrack {C(f_{i})\over f_{i}}\biggr\rbrack'}{\bf R}_{i}(t),
\end{eqnarray}
where $-\xi{\bf v}_{i}$ is a friction force and $U({\bf
r}_{1},...,{\bf r}_{N})=\sum_{i<j}u({\bf r}_{i}-{\bf r}_{j})$ is a
potential of interaction. The last term is a generalized stochastic
force. The usual white noise term ${\bf R}_{i}(t)$, satisfying
$\langle {\bf R}_{i}(t)\rangle={\bf 0}$ and $\langle
{R}_{a,i}(t){R}_{b,j}(t')\rangle=\delta_{ij}\delta_{ab}\delta(t-t')$,
where $a,b=1,...,D$ refer to the coordinates of space and
$i,j=1,...,N$ to the particles, is multiplied by a function that can depend
on the local (macroscopic) distribution function $f_{i}\equiv f({\bf
r}_{i},{\bf v}_{i},t)$. When
$C(f)=f \ln f$, which is related to the Boltzmann entropy (\ref{ex1}),
Eq. (\ref{a16}) reduces to the usual Langevin equations
\begin{equation}
\label{a16s}
{d{\bf r}_{i}\over dt}={\bf v}_{i},\quad
{d{\bf v}_{i}\over dt}=-\xi{\bf v}_{i}-\nabla_{i}U({\bf r}_{1},...,{\bf r}_{N})+\sqrt{2D}{\bf R}_{i}(t).
\end{equation}
This stochastic process describes a system of Brownian particles in
interaction. More generally, the term in front of ${\bf R}_{i}(t)$ in
Eq. (\ref{a16}) can be interpreted as a multiplicative noise since it
depends on ${\bf r},{\bf v}$. Note that it depends on ${\bf r},{\bf
v}$ through the distribution function $f({\bf r},{\bf
v},t)$. Therefore, there is a back-reaction from the macrodynamics,
leading to a situation of anomalous diffusion. In that context, the
generalized Kramers equation (\ref{gk2}) can be derived from the
N-body Fokker-Planck equation by using a Kramers-Moyal expansion and a
mean-field approximation
\cite{bcp2,bbgky}. The first term in Eq. (\ref{gk2}) is a generalized diffusion
(depending on the distribution function) and the second term is a
friction. The friction coefficient $\xi=D\beta$ satisfies a
generalized Einstein relation. Note that $D$ can depend on ${\bf r}$,
${\bf v}$ and $t$ without altering the general properties of the
equation. We can use this indetermination to write the generalized
Kramers equation in the alternative form
\begin{equation}
\label{gk3}
{\partial f\over\partial t}+Lf={\partial \over \partial {\bf v}}\cdot \biggl\lbrace D'\biggl\lbrack {\partial f\over\partial {\bf v}}+{\beta\over C''(f)}{\bf v}\biggr \rbrack \biggr\rbrace,
\end{equation}
which will have the same general properties as Eq. (\ref{gk2}). This
equation involves an ordinary diffusion and a nonlinear friction.
Equation (\ref{gk3}) can be deduced from Eq. (\ref{gk2}) by the
substitution $D'=DfC''(f)$. One of these two forms will be prefered
depending on the situation contemplated.

The generalized Kramers equation (\ref{gk2}) can also be obtained from a variational principle by maximizing the rate of free energy dissipation $\dot F$ at fixed mass and temperature \cite{gt}. Therefore, Eq. (\ref{gk2}) satisfies a canonical H-theorem $\dot F\le 0$, provided that $D\ge 0$. Indeed, an explicit calculation yields
\begin{equation}
\label{gk4}
\dot F=-\int {DT\over f}\biggl\lbrack fC''(f){\partial f\over\partial {\bf v}}+\beta f {\bf v}\biggr \rbrack^{2} d^{D}{\bf r}d^{D}{\bf v}\le 0.
\end{equation}
This shows that $F$ is the Lyapunov functional of the generalized
Kramers equation. Now, at equilibrium $\dot F=0$, so that according to
Eq. (\ref{gk4}),
\begin{equation}
\label{gk5}
{\partial C'(f)\over\partial {\bf v}}+\beta {\bf v}={\bf 0}.
\end{equation}
Integrating with respect to ${\bf v}$, we get
\begin{equation}
\label{gk6}
C'(f)=-\beta {v^{2}\over 2}+A({\bf r}).
\end{equation}
The cancellation of the advective term $Lf=0$ in Eq. (\ref{gk2})
combined with Eq. (\ref{gk6}) implies that $f=f(\epsilon)$ and
$\nabla A=-\beta\nabla\Phi$. Therefore, $A({\bf r})=-\beta\Phi({\bf
r})-\alpha$ and we recover Eq. (\ref{mep15}). This shows that a stationary
solution of Eq. (\ref{gk2}) extremizes the free energy $F$ at fixed
mass and temperature.  In addition, only {\it minima} of $F$ at fixed
$M$ and $T$ are linearly stable with respect to the generalized
Kramers equation (\ref{gk2}). Indeed, considering the linear
stability of a stationary solution of Eq. (\ref{gk2}),
we can derive the general relation
\begin{eqnarray}
\label{gk7}
2\lambda\delta^{2}{F}=\delta^{2}\dot F\le 0,
\end{eqnarray}
connecting the growth rate $\lambda$ of the perturbation $\delta f\sim
e^{\lambda t}$ to the second order variations of the free energy $F$
and the second order variations of the rate of free energy production
$\delta^{2}\dot F\le 0$ \cite{gt}.  Since the product
$\lambda\delta^{2}{F}$ is negative, we conclude that a stationary
solution of the generalized Kramers equation (\ref{gk2}) is linearly
stable ($\lambda<0$) if, and only if, it is a {\it minimum } of free
energy $F$ at fixed mass and temperature. This
aesthetic formula shows the equivalence between dynamical and
thermodynamical stability for the generalized Kramers
equation. Therefore, it only selects {\it minima} of $F$, not maxima
or saddle points.

\section*{8. Generalized Smoluchowski equation} We shall now introduce formally a relaxation equation associated with the minimization problem (\ref{fef16}). We write this equation in the form
\begin{equation}
\label{smol1}
{\partial \rho\over\partial t}=\nabla\cdot \biggl\lbrack
{1\over\xi}\rho \nabla \biggl ( {\delta F\over\delta \rho}\biggr
)\biggr\rbrack.
\end{equation}
Using the
expression of the free energy (\ref{fef15}), we obtain the generalized
Fokker-Planck equation
\begin{equation}
\label{smol2}
{\partial \rho\over\partial t}=\nabla\cdot\biggl\lbrack {1\over\xi}(\nabla p+\rho\nabla\Phi)\biggr\rbrack.
\end{equation}
Morphologically, Eq.  (\ref{smol2})
can be viewed as a generalized non-local Smoluchowski equation
\cite{risken}. It can be obtained from a variational principle by maximizing the rate of free energy dissipation $\dot F$ at fixed mass and temperature \cite{gt}. Therefore, Eq. (\ref{smol2}) satisfies a canonical H-theorem $\dot F\le 0$. Indeed, an explicit calculation yields
\begin{equation}
\label{smol3}
\dot F=-\int {1\over \xi\rho}(\nabla p+\rho\nabla\Phi)^{2}d^{D}{\bf r}\le 0.
\end{equation}
This shows that $F$ is the Lyapunov functional of the generalized
Smoluchowski equation. Now, at equilibrium $\dot F=0$, and we recover
the condition of hydrostatic balance (\ref{fef18}). It is also
possible to show \cite{gt} that only {\it minima} of free energy $F$
are linearly stable with respect to the generalized Smoluchowski
equation (\ref{smol2}).

Explicating the relation between the potential and the density,
the generalized Smoluchowski equation (\ref{smol2}) can be written
\begin{equation}
\label{smol4}
{\partial \rho\over\partial t}=\nabla\cdot\biggl\lbrace {1\over\xi}\biggl \lbrack p'(\rho)\nabla \rho+\rho\nabla\int u({\bf r}-{\bf r}')\rho({\bf r}',t)d^{D}{\bf r}'\biggr \rbrack\biggr\rbrace.
\end{equation}
The Lyapunov functional associated with this equation is the free energy
\begin{eqnarray}
{F}[\rho]=\int \rho \int_{0}^{\rho}{p(\rho')\over\rho'^{2}}d\rho'
d^D{\bf r}+{1\over
  2}\int\rho({\bf r},t)u({\bf r}-{\bf r}')\rho({\bf r}',t) d^{D}{\bf r}d^{D}{\bf r}'.
\label{smol5}
\end{eqnarray}
The stationary states are determined by the integro-differential
equation
\begin{equation}
\label{smol6} {p'(\rho)\over \rho}\nabla \rho=-\nabla\int \rho({\bf
r}')u({\bf r}-{\bf r}')d^{D}{\bf r}'.
\end{equation}
Equation (\ref{smol4}) generalizes the usual Smoluchowski equation in
two respects.  First, the coefficient of diffusion
${1\over\xi}p'(\rho)$ is an arbitrary function of the density $\rho$
associated with a generalized entropy functional (first term in
Eq. (\ref{smol5})). Second, this equation is non-local due to the
presence of an arbitrary binary potential of interaction $u({\bf
r}-{\bf r}')$ in the energy functional (second term in Eq.
(\ref{smol5})).  This equation possesses therefore a very rich
mathematical and physical structure.

We have started a systematic study of the generalized Smoluchowski
equation (\ref{smol4}) by considering first the gravitational
interaction in $D$ dimensions. The case of an isothermal equation of
state corresponding to the Boltzmann free energy has been considered
in \cite{crs,sire,post,tcoll}. It describes a system of
self-gravitating Brownian particles (see a review in \cite{revbd}). The
case of a quantum equation of state associated with the Fermi-Dirac
free energy has been considered in \cite{bf,bcp2}. It describes a
system of self-gravitating Brownian fermions. Finally the case of a
polytropic equation of state associated with Tsallis free energy has
been considered in
\cite{anomalous}. It describes a system of self-gravitating Langevin
particles experiencing anomalous diffusion.

\section*{9. The method of moments} The Smoluchowski equation can be
derived from the Kramers equation in a high friction limit
$\xi\rightarrow +\infty$, or equivalently for large times $t\gg
\xi^{-1}$. This can be shown easily by using a method of
moments. Integrating Eq. (\ref{gk2}) over velocity, we get the
continuity equation
\begin{eqnarray}
\label{mom1}
{\partial\rho\over\partial t}+\nabla\cdot  (\rho {\bf u})=0,
\end{eqnarray}
where ${\bf u}=(1/\rho)\int f{\bf v}d^{3}{\bf v}$ is the local velocity.
Multiplying Eq. (\ref{gk2}) by ${\bf v}$ and integrating over velocity, we get the momentum equation
\begin{equation}
\label{mom2}
{\partial\over\partial t}(\rho u_{i})+{\partial\over\partial x_{j}}(\rho u_{i}u_{j})+{\partial\over\partial x_{j}}P_{ij}+\rho{\partial\Phi\over\partial x_{i}}=-\int D\biggl \lbrack f C''(f){\partial f\over\partial v_{i}}+\beta f v_{i}\biggr \rbrack d^{3}{\bf v},
\end{equation}
where $P_{ij}=\int f w_{i}w_{j} d^{3}{\bf v}$ is the pressure tensor
and ${\bf w}={\bf v}-{\bf u}$ the relative velocity. Introducing the notation
$\phi(f)=\int^{f}xC''(x)dx$, the first term in the right hand side can
be rewritten $\partial
\phi(f)/\partial {\bf v}$ and, since it is a gradient of a function,
it vanishes by integration. We are left therefore with
\begin{eqnarray}
\label{mom3}
{\partial\over\partial t}(\rho u_{i})+{\partial\over\partial x_{j}}(\rho u_{i}u_{j})+{\partial\over\partial x_{j}}P_{ij}+\rho{\partial\Phi\over\partial x_{i}}=-\xi\rho  u_{i}.
\end{eqnarray}

In the high friction
limit $\xi=D\beta\rightarrow +\infty$, assuming $\beta$ of order
unity, the term in bracket in Eq. (\ref{gk2}) must vanish so that the
distribution function satisfies
\begin{eqnarray}
\label{mom4}
C'(f)=-\beta\biggl \lbrack {v^{2}\over 2}+\lambda({\bf r},t)\biggr \rbrack+O(\xi^{-1}),
\end{eqnarray}
where $\lambda({\bf r},t)$  is related to the
spatial density $\rho({\bf r},t)$ through the relation
\begin{eqnarray}
\label{mom5}
\rho=\int f d^{D}{\bf v}.
\end{eqnarray}
We find therefore that ${\bf u}=O(\xi^{-1})$ and $P_{ij}=p\delta_{ij}+O(\xi^{-1})$ where $p({\bf r},t)$ is the local pressure
\begin{eqnarray}
\label{mom6}
p={1\over D}\int f v^{2}d^{D}{\bf v},
\end{eqnarray}
determined by Eq. (\ref{mom4}). As in Sec. 4, the fluid is {\it barotropic},
i.e. $p({\bf r},t)=p\lbrack \rho({\bf r},t)\rbrack$. To first order in $\xi^{-1}$, the momentum equation (\ref{mom3}) implies that
\begin{eqnarray}
\label{mom7}
\rho {\bf u}=-{1\over\xi}(\nabla p+\rho\nabla\Phi)+O(\xi^{-2}).
\end{eqnarray}
Inserting the relation (\ref{mom7}) in the continuity equation
(\ref{mom1}), we get the generalized Smoluchowski equation
(\ref{smol2}).  The generalized Smoluchowski equation, as well as the
first order correction to the distribution function $f({\bf r},{\bf
v},t)$, can also be obtained from a formal Chapman-Enskog expansion
\cite{lemou}.

\section*{10. Generalized kinetic equations} The Kramers equation and the Smoluchowski equation have a canonical structure in which the temperature is fixed. We shall now introduce a generalized kinetic equation possessing a microcanonical structure in which the energy is fixed \cite{gt,gtkin}. This is the generalized Landau equation
\begin{eqnarray}
\label{landau1} {\partial f\over \partial t}+Lf ={\partial\over\partial
v^{\mu}}\int  d^{3}{\bf v}_{1}\ K^{\mu\nu} f f_{1}\biggl\lbrace
C''(f){\partial f\over\partial v^{\nu}}-C''(f_{1}){\partial
f_{1}\over\partial v_{1}^{\nu}} \biggr\rbrace,
\end{eqnarray}
with
\begin{equation}
\label{landau2}
K^{\mu\nu}={A\over u}\biggl (\delta^{\mu\nu}-{u^{\mu}u^{\nu}\over u^{2}}\biggr ),
\end{equation}
where ${\bf u}={\bf v}_{1}-{\bf v}$ is the relative velocity and $A$
is a constant.  We can also
consider the alternative form
\begin{eqnarray}
\label{landau3}
{\partial f\over \partial t}+Lf ={\partial\over\partial v^{\mu}}\int  d^{3}{\bf v}_{1}\ K^{\mu\nu} \biggl\lbrace
{1\over C''(f_{1})}{\partial f\over\partial v^{\nu}}-{1\over C''(f)}{\partial f_{1}\over\partial v_{1}^{\nu}}\biggr\rbrace.
\end{eqnarray}
The generalized Landau equation can be derived from a generalized
Boltzmann equation in a weak deflexion approximation. In turn, the
generalized Kramers equation (\ref{gk2}) can be derived from the
generalized Landau equation in a thermal bath approximation
\cite{gtkin}. The generalized Landau equation satisfies a H-theorem
($\dot S\ge 0$) for the generalized entropy (\ref{mep5}). The entropy
(\ref{mep5}) plays therefore the role of a Lyapunov functional. At
equilibrium, $\dot S=0$, and we obtain the distribution (\ref{mep8}).
In addition, it can be shown \cite{gtkin} that only {\it maxima} of
$S$ at fixed $M$ and $E$ are linearly stable with respect to the
generalized Landau equation.

Generalized kinetic equations such as (\ref{landau1}) appear when the
transition probabilities are different from the form that we would
naively expect due to the action of microscopic constraints (hidden
constraints). One particular case is when the particles are fermions.
In that case, the ``hidden constraints'' correspond to the Pauli
exclusion principle that prevents two particles with equal spin to
occupy the same phase space cell. This exclusion has been explained by
quantum mechanics. More generally, the ``hidden constraints'' may not
have necessarily a fundamental origin.

\section*{11. Particular examples} It can be of interest to discuss some
special cases explicitly. For the Boltzmann entropy (\ref{ex1}),
$C''(f)=1/f$ and Eq. (\ref{gk2}) reduces to the ordinary Kramers
equation
\begin{equation}
\label{pe1}
{\partial f\over\partial t}+L f={\partial \over \partial {\bf v}}\cdot \biggl\lbrack D\biggl ( {\partial f\over\partial {\bf v}}+\beta f{\bf v}\biggr ) \biggr\rbrack.
\end{equation}
The corresponding equation in physical space, obtained in the high
friction limit, is the ordinary Smoluchowski equation
\begin{eqnarray}
\label{pe2}
{\partial\rho\over\partial t}=\nabla \cdot \biggl\lbrack {1\over\xi}(T\nabla \rho+\rho\nabla\Phi)\biggr\rbrack.
\end{eqnarray}
Finally, Eq. (\ref{landau1}) leads to the ordinary Landau equation
\begin{eqnarray}
\label{blandau1}{\partial f\over\partial t}+L f={\partial\over\partial
v^{\mu}}\int  d^{3}{\bf v}_{1}\ K^{\mu\nu}\biggl ( f_{1}{\partial
f\over\partial v^{\nu}}-f{\partial f_{1}\over\partial
v_{1}^{\nu}}\biggr ).
\end{eqnarray}

For the Fermi-Dirac entropy (\ref{ex6}), $C''(f)=1/f(\eta_{0}-f)$. In
order to avoid the divergence of the term $f C''(f)$ as $f\rightarrow
\eta_{0}$, it is appropriate to consider the alternative form
(\ref{gk3}) of the generalized Kramers equation. This yields
\begin{equation}
\label{pe3}
{\partial f\over\partial t}+L f={\partial \over \partial {\bf v}}\cdot \biggl\lbrace D'\biggl \lbrack {\partial f\over\partial {\bf v}}+\beta f(\eta_{0}-f){\bf v}\biggr \rbrack \biggr\rbrace.
\end{equation}
The corresponding equation in physical space is given by
Eq. (\ref{smol2}) with the equation of state (\ref{ex10}). In fact,
when the Kramers equation is written in the form (\ref{gk3}), the
coefficient in factor of the diffusion current in Eq. (\ref{smol2}) is
{\it not} $1/\xi$ but is more complex. We refer to \cite{lemou} for a
detailed discussion of this subtle point. Finally, Eq. (\ref{landau3})
leads to the fermionic Landau equation
\begin{eqnarray}
\label{blandau2} {\partial f\over\partial t}+L f={\partial\over\partial v^{\mu}}\int
d^{3}{\bf v}_{1}\ K^{\mu\nu} \biggl\lbrace f_1 (\eta_{0}- f_1)
{\partial f\over\partial v^{\nu}}-f (\eta_{0}- f){\partial
f_{1}\over\partial v_{1}^{\nu}}\biggr\rbrace.
\end{eqnarray}

For the Tsallis entropy (\ref{ex11}), $C''(f)=q f^{q-2}$ and
Eq. (\ref{gk2}) has the form of a nonlinear Kramers equation
\begin{equation}
\label{pe4}
{\partial f\over\partial t}+L f={\partial \over \partial {\bf v}}\biggl\lbrack D\biggl ( {\partial f^{q}\over\partial {\bf v}}+\beta f {\bf v}\biggr ) \biggr\rbrack .
\end{equation}
The corresponding equation in physical space is the nonlinear Smoluchowski equation
\begin{eqnarray}
\label{pe5}
{\partial\rho\over\partial t}=\nabla \biggl\lbrack {1\over\xi}(K\nabla \rho^{\gamma}+\rho\nabla\Phi)\biggr\rbrack.
\end{eqnarray}
Finally, Eq. (\ref{landau1}) leads to the $q$-Landau equation
\begin{eqnarray}
\label{s6} {\partial f\over\partial t}+L f={\partial\over\partial
v^{\mu}}\int  d^{3}{\bf v}_{1}\ K^{\mu\nu}\biggl (f_1 {\partial
f^q\over\partial v^{\nu}}-f {\partial f_{1}^q\over\partial
v_{1}^{\nu}}\biggr ).
\end{eqnarray}

\section*{12. Maximum entropy principle in physical space} We now consider a system of $N$
particles in interaction with total mass
\begin{equation}
\label{ps1} M=\int \rho({\bf r},t) \ d^{D}{\bf r},
\end{equation}
and energy
\begin{equation}
\label{ps2} E={1\over 2}\int \rho \Phi \ d^{D}{\bf  r},
\end{equation}
where the potential $\Phi({\bf r},t)$ is related to the density
$\rho({\bf r},t)$ by a relation of the form
\begin{equation}
\label{ps3}\Phi({\bf r},t)=\int \rho({\bf r}',t)u({\bf r}-{\bf
r'})d^{D}{\bf r}'.
\end{equation}

We introduce a generalized entropy in physical space
\begin{equation}
\label{ps4} S=-\int C(\rho)\ d^{D}{\bf r},
\end{equation}
where $C(\rho)$ is a convex function, i.e. $C''(\rho)> 0$. We are
interested by the distribution $\rho({\bf r})$ which maximizes the
generalized entropy (\ref{ps4}) at fixed mass and energy, i.e.
\begin{equation}
\label{ps5} {\rm Max}\quad S[\rho]\quad | \quad E[\rho]=E,\
M[\rho]=M.
\end{equation}
Since the energy is fixed, we shall associate this maximization
problem to a microcanonical description. Introducing Lagrange
multipliers and writing the variational principle in the form
\begin{equation}
\label{ps6} \delta S-\beta\delta E-\alpha\delta M=0,
\end{equation}
we find that the critical points of entropy at fixed mass and
energy are given by
\begin{equation}
\label{ps7} C'(\rho)=-\beta\Phi-\alpha.
\end{equation}
 The Lagrange multipliers $\beta=1/T$
and $\alpha$ can be interpreted as a generalized inverse
temperature and a generalized chemical potential, respectively.
Equation (\ref{ps7}) can be written equivalently as
\begin{equation}
\label{ps8} \rho=F(\beta\Phi+\alpha),
\end{equation}
where $F(x)=(C')^{-1}(-x)$. From the identity
\begin{equation}
\label{ps9} \rho'(\Phi)=-\beta/C''(\rho),
\end{equation}
resulting from Eq. (\ref{ps7}), $\rho(\Phi)$ is a monotonically
decreasing function of $\Phi$ if $\beta>0$ and a monotonically
increasing function of $\Phi$ if $\beta<0$. Explicating the
relation between the potential and the density, we find that the
equilibrium distribution is determined by the integro-differential
equation
\begin{equation}
\label{ps10} C'(\rho)=-\beta\int \rho({\bf r}')u({\bf r}-{\bf
r}')d^{D}{\bf r}'-\alpha.
\end{equation}

We now introduce the generalized free energy
\begin{equation}
\label{ps11} J[\rho]=S[\rho]-\beta E[\rho],
\end{equation}
associated with the functionals (\ref{ps2}) and (\ref{ps4}). We are
interested by the density $\rho({\bf r})$ which maximizes the
generalized free energy (\ref{ps11}) at fixed mass and temperature,
i.e.
\begin{equation}
\label{ps12} {\rm Max}\quad J[\rho]=S[\rho]-\beta E[\rho]\quad |
\quad M[\rho]=M.
\end{equation}
Since the temperature is given, we shall associate this
maximization problem to a canonical description. Introducing
Lagrange multipliers and writing the variational principle in the
form
\begin{equation}
\label{ps13} \delta J-\alpha\delta M=0,
\end{equation}
we find that the critical points of free energy at fixed mass and
temperature are given by
\begin{equation}
\label{ps14} C'(\rho)=-\beta\Phi-\alpha,
\end{equation}
as in the microcanonical description.

\section*{13. Stability conditions} In the microcanonical situation, 
we must select {\it maxima} of $S[\rho]$ at fixed mass and energy. The
condition that $\rho$ is a {maximum} of $S$ at fixed mass and energy is
equivalent to the condition that $\delta^{2}{J}\equiv
\delta^{2}{S}-\beta\delta^{2}E$ is negative for all perturbations
that conserve mass and energy to first order. This condition can
be written
\begin{eqnarray}
\label{sc1}
\delta^{2}J=-\int C''(\rho){(\delta \rho)^{2}\over 2}d^{D}{\bf  r}
-{1\over 2}\beta\int \delta\rho\delta\Phi d^{D}{\bf  r}\le 0,\nonumber\\
\forall\ \delta \rho \mid\ \delta E=\delta M=0.\qquad\qquad \qquad
\end{eqnarray}
The condition of stability in the canonical situation requires
that $\rho$ is a {\it maximum} of $J[\rho]$ at fixed mass and
temperature. This is equivalent to the condition that
$\delta^{2}{J}$ is negative for all perturbations that conserve
mass. This can be written
\begin{eqnarray}
\label{sc2}
\delta^{2}J=-\int C''(\rho){(\delta \rho)^{2}\over 2}d^{D}{\bf  r}
-{1\over 2}\beta\int \delta\rho\delta\Phi d^{D}{\bf  r}\le 0,\nonumber\\
\forall\ \delta \rho \mid\ \delta M=0.\qquad\qquad \qquad
\end{eqnarray}

Using Eq. (\ref{ps9}), the functional arising in these stability criteria
can be expressed as
\begin{eqnarray}
\label{sc3} \delta^{2}J={1\over 2}\beta\biggl\lbrace \int {(\delta
\rho)^{2}\over \rho'(\Phi)}d^{D}{\bf  r}-\int \delta\rho\delta\Phi
d^{D}{\bf  r}\biggr\rbrace.
\end{eqnarray}
 The discussion on the inequivalence of the  microcanonical and canonical descriptions is the same as the one given in Sec. 3.

\section*{14. Examples of entropy functionals} The Boltzmann entropy in physical space
\begin{equation}
\label{e1} S_{B}[\rho]=-\int \rho\ln \rho d^{D}{\bf r},
\end{equation}
leads to the isothermal distribution
\begin{equation}
\label{e2} \rho=A e^{-\beta\Phi}.
\end{equation}

The Fermi-Dirac entropy in physical space
\begin{equation}
\label{e3} S_{FD}[\rho]=-\int \biggl\lbrace {\rho\over
\sigma_{0}}\ln {\rho\over \sigma_{0}}+\biggl (1-{\rho\over
\sigma_{0}}\biggr )\ln \biggl (1-{\rho\over \sigma_{0}}\biggr )
\biggr\rbrace d^{D}{\bf r},
\end{equation}
leads to the Fermi-Dirac distribution
\begin{equation}
\label{e4} \rho={\sigma_{0}\over 1+\lambda
e^{\beta\sigma_{0}\Phi}}.
\end{equation}
This distribution satisfies the constraint $\rho\le \sigma_{0}$ which
puts an upper bound on the local density of particles. It usually
arises when one considers finite size effects or when one introduces a
lattice model in physical space preventing two particles to occupy the
same site.  The isothermal distribution (\ref{e2}) is recovered in the
dilute limit $\rho\ll \sigma_{0}$.

The Tsallis entropy
\begin{equation}
\label{e5} S_{q}[\rho]=-{1\over q-1}\int (\rho^{q}-\rho) d^{D}{\bf r},
\end{equation}
where $q$ is a real number, leads to the polytropic distribution
\begin{equation}
\label{e6} \rho=\biggl\lbrack \mu-{(q-1)\beta\over q}\Phi\biggr\rbrack^{n},   \qquad n={1\over q-1}.
\end{equation}
The isothermal distribution 
(\ref{e2}) is recovered for $q\rightarrow 1$, i.e. $n\rightarrow +\infty$. 

\section*{15. Physical applications}The maximization problems discussed previously can have various physical applications (not necessarily related to thermodynamics) that we briefly mention.

{\it (i) Statistical mechanics of point vortices}: The variational
problems (\ref{ps5}) and (\ref{ps12}) arise in the statistical
mechanics of 2D point vortices provided that we make the
correspondance $\rho\leftrightarrow\omega$ between the density and the
vorticity and the correspondance $\Phi\leftrightarrow \psi$ between
the potential and the stream function. In that context, the
maximization problem (\ref{ps5}) associated with the Boltzmann entropy
(\ref{e1}) determines the statistical equilibrium state (most probable
distribution) of a cloud of point vortices \cite{jm,houches}.

{\it (ii) Two-dimensional Euler equation}: The maximization problem
(\ref{ps5}) determines nonlinearly dynamically stable 
stationary solutions of the 2D Euler-Poisson system
\begin{equation}
\label{paw1} {\partial \omega\over\partial t}+{\bf
u}\cdot\nabla\omega=0,
\end{equation}
\begin{equation}
\label{paw2} \Delta\psi=-\omega,
\end{equation}
where ${\bf u}=-{\bf z}\times\nabla\psi$ is the velocity field. The 2D
Euler equation describes the evolution of the density distribution of
point vortices in the ``collisionless'' regime (Vlasov regime) before
correlations (collisions) have developed \cite{kin}. It also describes
the inviscid evolution of continuous vorticity flows in
two-dimensional hydrodynamics. In that context, $S[\omega]$ is called
a H-function. Boltzmann and Tsallis functionals are particular
H-functions associated with isothermal and polytropic vortices
\cite{gt}. Note that the criterion of nonlinear dynamical stability
(\ref{sc1}) is more refined that the stability criterion (\ref{sc2})
which is itself more refined than the Arnol'd theorems \cite{ellis,bouchet,gt}.

{\it (iii) Generalized thermodynamics}: The variational principles
(\ref{ps5}) and (\ref{ps12}) determine the generalized
thermodynamical stability of complex systems exhibiting anomalous
diffusion in physical space. Anomalous diffusion can have various
origins: it can can be due to the existence of traps, of a lattice
preventing particles to reach occupied sites, or any other constraints
that are often difficult to formalize. This is what we call ``hidden
constraints'' \cite{gtkin}. These systems can be described by {\it
effective} equations resembling generalized Fokker-Planck equations
\cite{gt}.  The variational principles (\ref{ps5}) and (\ref{ps12})
determine the linear dynamical stability of stationary solutions of
these equations \cite{gt}.

\section*{16. Generalized drift-diffusion equation} We shall  introduce formally a relaxation equation associated with the maximization problem (\ref{ps12}). We write this equation in the form 
\begin{equation}
\label{dd1} {\partial \rho\over\partial t}=-\nabla\cdot \biggl\lbrack
D\rho \nabla \biggl ({\delta J\over\delta \rho}\biggr )\biggr\rbrack,
\end{equation}
where $\delta/\delta \rho$ denotes the functional derivative. By
construction, this equation conserves mass since the right hand side
is the divergence of a current $-{\bf J}_{\rho}$. Using the expression
of the free energy (\ref{ps11}) with Eqs. (\ref{ps2}) and (\ref{ps4}),
we obtain the generalized drift-diffusion equation
\begin{equation}
\label{dd2} {\partial \rho\over\partial t}=\nabla\cdot \biggl\lbrace
D\biggl\lbrack \rho C''(\rho)\nabla\rho+\beta \rho\nabla\Phi\biggr
\rbrack \biggr\rbrace.
\end{equation}
Morphologically, this equation can be viewed as a generalized
non-local Smoluchowski equation. It describes the dynamics
(in physical space) of a system of Langevin particles in interaction
governed by a generalized class of stochastic processes \cite{gt,bbgky}:
\begin{equation}
\label{a4}
{d{\bf r}_{i}\over dt}=-\xi\nabla_{i}U({\bf r}_{1},...,{\bf r}_{N})+\sqrt{2D\rho_{i}\biggl\lbrack {C(\rho_{i})\over\rho_{i}}\biggr\rbrack'}{\bf R}_{i}(t),
\end{equation}
where $\rho_{i}\equiv \rho({\bf r}_{i},t)$ and ${\bf R}_{i}(t)$ is a
white noise satisfying $\langle {\bf R}_{i}(t)\rangle={\bf 0}$ and
$\langle
{R}_{a,i}(t){R}_{b,j}(t')\rangle=\delta_{ij}\delta_{ab}\delta(t-t')$,
where $a,b=1,...,D$ refer to the coordinates of space and
$i,j=1,...,N$ to the particles. The particles interact via the
potential $U({\bf r}_{1},...,{\bf r}_{N})=\sum_{i<j}u({\bf r}_{i}-{\bf
r}_{j})$ and $C(\rho)$ is an arbitrary convex function. When
$C(\rho)=\rho\ln\rho$, which is related to the Boltzmann entropy (\ref{e1}),
Eq. (\ref{a4}) reduces to the usual Langevin equations
\begin{equation}
\label{a5}
{d{\bf r}_{i}\over dt}=-\xi\nabla_{i}U({\bf r}_{1},...,{\bf r}_{N})+\sqrt{2D}{\bf R}_{i}(t).
\end{equation}
This stochastic process describes a system of Brownian particles in
interaction. The term in front of ${\bf R}_{i}(t)$ in Eq. (\ref{a4})
can be interpreted as a multiplicative noise since it depends on the
position ${\bf r}$. Note that it depends on ${\bf r}$ through the
density $\rho({\bf r},t)$. Therefore, there is a back-reaction from
the macrodynamics. In that context, the generalized Smoluchowski
equation (\ref{dd2}) can be obtained from the $N$-body Fokker-Planck
equation by using a Kramers-Moyal expansion and a meanfield
approximation \cite{bbgky,bcp2}. The first term in Eq. (\ref{dd2}) is
a generalized diffusion (depending on the density) and the second term
is a drift. The drift coefficient $\xi=D\beta$ satisfies a generalized
Einstein relation. Note that $D$ can depend on ${\bf r}$ and $t$
without altering the general properties of the equation. We can use
this indetermination to write the generalized drift-diffusion equation
in the alternative form
\begin{equation}
\label{dd3} {\partial \rho\over\partial t}=\nabla\cdot \biggl\lbrace
D'\biggl\lbrack \nabla\rho+{\beta\over C''(\rho)}\nabla\Phi\biggr
\rbrack \biggr\rbrace,
\end{equation}
which will have the same general properties as Eq. (\ref{dd2}).
This equation involves an ordinary diffusion and a nonlinear
drift. Equation (\ref{dd3}) can be deduced from Eq. (\ref{dd2}) by
the substitution $D'=D\rho C''(\rho)$. One of these two forms will
be prefered depending on the situation contemplated.

The generalized drift-diffusion equation (\ref{dd2}) can also be
obtained from a variational principle by maximizing the rate of free
energy production $\dot J$ at fixed mass and temperature
\cite{gt}. Therefore, Eq.  (\ref{dd2}) satisfies a canonical H-theorem $\dot
J\ge 0$. An explicit calculation yields
\begin{equation}
\label{dd4} \dot J=\int {D\over \rho}\biggl\lbrack \rho
C''(\rho)\nabla\rho+\beta \rho \nabla\Phi\biggr \rbrack^{2}
d^{D}{\bf r}\ge 0.
\end{equation}
If $\rho\ge 0$, this inequality is true provided that $D>0$. If $\rho$
can take both signs (which depends on the initial conditions), it is
more convenient to consider Eq. (\ref{dd3}).  In that case, the
inequality $\dot J\ge 0$ is true provided that $D'>0$. This shows that
$J$ is the Lyapunov functional associated with the generalized
drift-diffusion equation (\ref{dd2}).

Now, at equilibrium $\dot J=0$, so that according to Eq. (\ref{dd4}),
\begin{equation}
\label{dd5} \nabla C'(\rho)+\beta \nabla\Phi={\bf 0}.
\end{equation}
Integrating, we get
\begin{equation}
\label{dd6} C'(\rho)=-\beta \Phi-\alpha,
\end{equation}
and we recover Eq. (\ref{ps14}). Therefore, a stationary solution of
Eq. (\ref{dd2}) extremizes the free energy $J$ at fixed mass and
temperature.  In addition, it is shown in \cite{gt} that only {\it
maxima} of $J$ at fixed $M$ and $T$ are linearly stable with respect
to the generalized Fokker-Planck equation (\ref{dd2}).

Explicating the relation between the potential and the density,
the generalized drift-diffusion equation (\ref{dd3}) can be written
\begin{equation}
\label{dd7} {\partial \rho\over\partial t}=\nabla\cdot\biggl\lbrace
D\biggl \lbrack \nabla \rho+{\beta\over C''(\rho)}\nabla\int
u({\bf r}-{\bf r}')\rho({\bf r}',t)d^{D}{\bf r}'\biggr
\rbrack\biggr\rbrace,
\end{equation}
with the free energy
\begin{equation}
\label{dd7b} J=-\int C(\rho)\ d^{D}{\bf r}-{1\over 2}\beta\int \rho\Phi \ d^{D}{\bf r}.
\end{equation}
This equation generalizes many drift-diffusion equations introduced in
the literature (we refer to \cite{biler} for a connexion with
mathematical works and for a detailed list of references). It would be
of interest to investigate its properties by remaining as general as
possible, i.e. without specifying the function $C(\rho)$ and the
binary potential of interaction $u({\bf r}-{\bf r}')$ explicitly. For
example, which conditions must satisfy $C(\rho)$ and $u({\bf r}-{\bf
r}')$ to generate blow-up solutions? Can we regroup the functionals
$S[\rho]$ is ``classes of equivalence'' with the heuristic idea that
functionals of the same ``class'' will yield ``similar'' results?
Which functionals $S[\rho]$ lead to confined solutions with a compact
support, such as polytropic distributions associated with the Tsallis
functional \cite{anomalous}? These are interesting mathematical
problems which could be tackled in relation with Eq. (\ref{dd7}).

\section*{17. Relation to Cahn-Hilliard equations.} If we now consider 
the case of short-range interactions, it is
possible to expand the potential
\begin{equation}
\label{ch1}
\Phi({\bf r},t) = \int u({\bf r}')\rho({\bf r}+{\bf r'}) d^D{\bf
  r}'
\end{equation}
in Taylor series for ${\bf r}'\to {\bf 0}$. Introducing the notations
\begin{equation}
\label{ch2}
a=\int u(|{\bf x}|) d^{D}{\bf x} \;\;\mbox{ and }\;\; b=
{1\over D} \int u(|{\bf x}|)x^2 d^{D}{\bf x}\,,
\end{equation}
we obtain to second order
\begin{equation}
\label{ch3}
\Phi({\bf r},t) = a\rho({\bf r},t) + {b\over 2} \Delta\rho({\bf
  r},t)\,.
\end{equation}
In that limit, the free energy takes the form
\begin{equation}
\label{ch4}
J[\rho] = {1\over 2}\beta b \int \left\{ {(\nabla\rho)^2\over 2} +
  V(\rho) \right\} d^D{\bf r}\,,
\end{equation}
where we have set $V(\rho) = -(2/b\beta)C(\rho) - (a/b)\rho^2$.  This
is the usual expression of the Landau free energy. In general, $\beta
b$ is negative so that we have to minimize the functional integral. For
systems with short-range interactions, the conservative equation
(\ref{dd7}) becomes
\begin{equation}
\label{ch5}
{\partial\rho\over\partial t} = \nabla\cdot
\biggl\lbrace {b\xi\over 2}\rho\nabla \left( \Delta\rho - V'(\rho) \right)
\biggr\rbrace\,.
\end{equation}
This is the Cahn-Hilliard equation which has been extensively studied
in the theory of phase ordering kinetics. Its stationary solutions
describe ``domain walls''. We can view therefore Eq.~(\ref{dd7})
as a generalization of the Cahn-Hilliard equation to the case of
systems with long-range interactions. Therefore, its general study is of 
great mathematical and physical interest.

\section*{18. Physical applications}  The generalized Fokker-Planck equations discussed previously and in \cite{gt,gtkin} can have various physical applications (not necessarily related to thermodynamics) that we briefly mention.

{\it (i) Numerical algorithms}: The relaxation equations presented in
Secs. 7, 8 and 16, and in Refs. \cite{gt,gtkin} can be used as
numerical algorithms to solve the variational problems (\ref{mep6}),
(\ref{mep13}), (\ref{fef16}), (\ref{ps5}) and (\ref{ps12}).  This is
of great practical interest because it is in general difficult to
solve the integrodifferential equations (\ref{mep11}), (\ref{fef21})
and (\ref{ps10}) directly and be sure that the solution is a maximum
of $S$ or $J$. These relaxation equations can thus be used as
numerical algorithms to construct thermodynamically stable equilibrium
states of Hamiltonian systems with long-range interactions (see Sec. 6
(i) and 15 (i)) as well as nonlinearly dynamically stable stationary
solutions of the Vlasov, Euler-Jeans and 2D Euler equations (see
Secs. 6 (ii), (iii) and 15 (ii)).

{\it (ii) Violent relaxation:} The relaxation equations introduced in
\cite{rsmepp,csr,gt,gtkin} provide a small-scale parameterization of
the Vlasov-Poisson and 2D Euler-Poisson systems. They describe the
convergence of the flow, on a coarse-grained scale, towards a
metaequilibrium state corresponding to a galaxy in astrophysics or a
large-scale vortex (e.g., Jupiter's great red spot) in 2D
hydrodynamics.  The theory of violent relaxation is discussed in,
e.g., \cite{vr,csr,dubrovnik,houches}.

{\it (iii) Statistical mechanics:} In the isothermal case, the
nonlocal Kramers and Smoluchowski equations describe the evolution of
the distribution function and density of a gas of Brownian particles
in interaction \cite{revbd,bcp2}. They can be seen, therefore, as the
canonical counterpart of the nonlocal Vlasov, Boltzmann and Landau
equations describing the evolution of the distribution function of a
Hamiltonian system of particles in interaction, for which a
microcanonical description holds
\cite{bbgky}. Since statistical ensembles are not equivalent for
systems with long-range interactions, it is of conceptual interest to
compare the microcanonical and canonical descriptions to see the
analogies and the differences \cite{crs}.

{\it (iii) Biological colonies:} Non-local drift-diffusion equations
also occur in biology, in connexion with the chemotactic aggregation of
bacterial populations. A model of chemotactic aggregation has been
proposed by Keller \& Segel
\cite{ks}. In some approximation \cite{jager}, their equations reduce to the
Smoluchowski-Poisson system as for self-gravitating Brownian particles
\cite{crs,revbd}. This analogy is developed in \cite{bcp2}. In a more general context, the diffusion coefficient or the
chemotactic drift can depend on the density. This can take into
account anomalous diffusion or finite size effects preventing
unphysical blow-up of the bacterial concentration. A simple
regularization of the Smoluchowski-Poisson system is provided by the
equation
\begin{equation}
\label{dd2bis} {\partial \rho\over\partial t}=\nabla\cdot \biggl\lbrace
D\biggl\lbrack \nabla\rho+\beta \rho (1-\rho/\sigma)\nabla\Phi\biggr
\rbrack \biggr\rbrace,
\end{equation}
which is associated with the Fermi-Dirac entropy (\ref{e3}) in
physical space. This equation respects the constraint $\rho\le \sigma$
at any time and therefore prevents blow-up. By rescaling the diffusion
coefficient $D({\bf r},t)$ appropriately, this equation can be put in
the form (\ref{smol2}) with an effective pressure $p=T\ln
(1-\rho/\sigma)$. A more general model of chemotaxis is provided by
the drift-diffusion equation (\ref{dd7}) which can take into account
anomalous diffusion and finite size effects of various forms. As we
have seen, this equation is associated with an effective
thermodynamical formalism. An even more general model is represented
by the non-Markovian equation
\begin{equation}
\label{fin1} {\partial \rho\over\partial t}=\nabla\cdot\biggl\lbrace
D\biggl \lbrack \nabla \rho+{\beta\over C''(\rho)}\nabla\int\int_{0}^{t}
u({\bf r}-{\bf r}',t-t')\rho({\bf r}',t')d^{D}{\bf r}'dt'\biggr
\rbrack\biggr\rbrace.
\end{equation}
This equation can take into account delay effects that are relevant in
the problem of chemotaxis (indeed, the equation satisfied by $\Phi$ is
of the form $\epsilon\partial_{t}\Phi=\Delta\Phi-\lambda\rho-\mu\Phi$
\cite{ks}). We note that Eq. (\ref{fin1}) does not admit a Lyapunov
functional (or a free energy) so that the effective thermodynamical
formalism developed previously breaks up. It would be of interest to
see how these ideas can be generalized to that context.

{\it (iii) Generalized thermodynamics:} The relaxation equations
presented in Secs. 7, 8 and 16, and in Refs. \cite{gt,gtkin} can be
considered as {\it effective} kinetic equations which may be useful to
model complex systems. As indicated previously, they can arise when
the system is subject to microscopic (hidden) constraints that change
the form of the transition probabilities. The consequence is that the
accessible microstates are not equiprobable, resulting in new forms of
entropic functionals. An explicit and fundamental example is played by
the Pauli exclusion principle in quantum mechanics which leads to the
Fermi-Dirac entropy instead of the Boltzmann entropy. We have also
exhibited a generalized stochastic process (\ref{a16})-(\ref{a4})
which leads to generalized Fokker-Planck equations of various
forms. These stochastic equations form just a particular example of
processes generating anomalous diffusion and complex phase space
structure. More general and more realistic microscopic processes could
also be considered and studied.  However, since the generalized
kinetic equations (\ref{gk2}), (\ref{smol2}) and (\ref{dd2}) can be
obtained from arguments of a very wide scope, such as the Maximum
Entropy (Free Energy) Production (Dissipation) Principle \cite{gt} for
example, we believe that they have a relatively fundamental and
universal character and that they will be obtained from a large class
of microscopic processes.

\section*{19. Conclusion}  In this paper, we have developed 
a generalized thermodynamical formalism and corresponding kinetic
theories. We believe that generalized thermodynamics is relevant for
the physics of complex systems when ``hidden constraints'' are in
action \cite{gtkin}. These microscopic constraints imply that the
microstates with given mass and energy are {\it not} equiprobable,
contrary to what is usually {\it postulated} in thermodynamics.  We
can either use the Boltzmann entropy and try to take into account
these additional microscopic constraints, or consider only the usual
constraints (mass and energy) and change the form of  entropy
\cite{gtkin}. This introduces some indetermination that is
encapsulated in the $q$ parameter of Tsallis or more generally in a
function $C(f)$. We emphasize that this indetermination is {\it
intrinsic} to the problem and not a flaw of our approach. It occurs
because we do not have a complete knowledge of the system's
dynamics. However, we have suggested \cite{gt} that generalized
entropies can be regrouped in ``classes of equivalence'' and that, for
a given system, a class is more appropriate than another. This extends
the attempt of Tsallis and co-workers to try to find the correct value
of $q$ corresponding to a given (complex) system. We argue that, more
than a scalar $q$, a full function $C(f)$ must be considered in
general.

On the other hand, the maximization problems $\lbrace {\rm Max}\ S\ |\
E, M\ {\rm fixed}\rbrace$ and $\lbrace{\rm Min}\ F=E-TS\ |\ M\ {\rm
fixed}\rbrace$ can arise in different situations that are not
necessarily related to thermodynamics. For example, the maximization
of a $H$ function at fixed mass and energy determines nonlinearly
dynamically stable stationary solutions of the Vlasov-Poisson system
in astrophysics. In that case, this maximization problem is related to
dynamics, not thermodynamics. However, in order to investigate this
dynamical problem, it can be useful to develop a {\it thermodynamical
analogy} and use a similar vocabulary. We have given other examples
(in fluid mechanics, biology,...) where this thermodynamical
analogy could be developed.

%%%%%enter the widest reference label as the first parameter of ref%%%%%
%\references{Nov}
%{\item{[6]} D. Beck,
%{\it Introduction to Dynamical Systems}\/,
%Vol.~2, Progr.~Math.~54, Birkh\"auser, Basel, 1978.
%
%\item{[7]}  R. Hill and A. Dow,
%{\it An index formula}\/,
%J.~Differential Equations 15 (1982), 197--211.
%
%\item{[8]}  J. Kowalski,
%{\it Some remarks on $J(X)$},  in: Algebra
%and Analysis (Edmonton, 1973),   E.~Brook (ed.),
%Lecture Notes in Math. 867, Springer, Berlin, 1974, 115--124.
%
%\item{[Nov]} A. S. Novikov,
%{\it An existence theorem for planar
%graphs}\/,  preprint, Moscow University, 1980 (in Russian).
%
%}


\begin{thebibliography}{99}

\bibitem{dauxois}  {\small  {\it Dynamics and thermodynamics of systems with long range interactions}, edited by Dauxois, T, Ruffo, S., Arimondo, E. and  Wilkens, M. Lecture Notes in Physics, Springer (2002).}

\bibitem{houches}  {\small  P.H. Chavanis, in {\it Dynamics and thermodynamics of systems with long range interactions}, edited by Dauxois, T, Ruffo, S., Arimondo, E. and  Wilkens, M. Lecture Notes in Physics, Springer (2002) [cond-mat/0212223].}

\bibitem{gt}  {\small P.H. Chavanis,  Phys. Rev. E. {\bf 68}, 036108 (2003).}

\bibitem{bcp2} {\small P.H. Chavanis, M. Ribot, C. Rosier and C. Sire, Banach Center Publications, this issue [cond-mat/0407386]. }

\bibitem{bbgky} {\small P.H. Chavanis, in preparation. }

\bibitem{gtkin}  {\small P.H. Chavanis,  Physica A  {\bf 332}, 89 (2004).}

\bibitem{lemou}  {\small  P.H. Chavanis, P. Lauren\c cot and M. Lemou, Physica A, in press [cond-mat/0403610].  }

\bibitem{tremaine}  {\small S. Tremaine, M. H\'enon and D. Lynden-Bell,  Mon. Not. R. astr. Soc. {\bf 219} 285 (1986).}

\bibitem{pt}  {\small P.H. Chavanis, Phys. Rev. E {\bf 65}, 056123 (2002).}

\bibitem{katz}  {\small J. Katz,  Mon. Not. R. astr. Soc.
{\bf 183}, 765 (1978).}

\bibitem{bb}  {\small F. Bouchet and J. Barr\'e, submitted to J. Stat. Phys. [cond-mat/0303307].}

\bibitem{tsallis}  {\small C. Tsallis,  J. Stat. Phys. {\bf 52}, 479 (1988).}

\bibitem{anomalous}  {\small  P.H. Chavanis and C. Sire, Phys. Rev. E {\bf 69}, 016116 (2004).   }

\bibitem{sire}  {\small C. Sire and P.H. Chavanis, Phys. Rev. E {\bf  66}, 046133 (2002).}

\bibitem{fermiD}  {\small P.H. Chavanis, Phys. Rev. E {\bf 69}, 066126 (2004).}

\bibitem{lbw}  {\small D. Lynden-Bell and R. Wood, Mon. Not. R. astr. Soc.
{\bf 138}, 495 (1968).}

\bibitem{paddy}  {\small T. Padmanabhan,  Phys. Rep.   {\bf 188}, 285 (1990).}

\bibitem{aa}  {\small P.H. Chavanis,  Astron. Astrophys. {\bf 381}, 340 (2002).}


\bibitem{crs}  {\small P.H. Chavanis, C. Rosier and C. Sire, Phys. Rev. E {\bf  66}, 036105 (2002).}

\bibitem{revbd}  {\small C. Sire and P.H. Chavanis, Banach Center Publications, this issue [cond-mat/0407397].}

\bibitem{grand}  {\small P.H. Chavanis, Astron. Astrophys. {\bf 401}, 15 (2003).}


\bibitem{vr}  {\small D. Lynden-Bell, MNRAS {\bf 136}, 101 (1967).}

\bibitem{csr}  {\small P.H. Chavanis, J. Sommeria and
R. Robert, Astrophys.
J. {\bf 471}, 385 (1996).}

\bibitem{bt}  {\small J. Binney and S. Tremaine, Galactic Dynamics, Princeton Series in Astrophysics (1987).}

\bibitem{risken}  {\small H. Risken, {The Fokker-Planck equation} (Springer, 1989).}

\bibitem{post}  {\small  C. Sire and P.H. Chavanis, Phys. Rev. E {\bf 69}, 066109 (2004).}

\bibitem{tcoll}  {\small  P.H. Chavanis and C. Sire, Phys. Rev. E, in press [cond-mat/0402227].}

\bibitem{bf} {\small P.H. Chavanis, M. Ribot, C. Rosier and C. Sire, in preparation. }

\bibitem{jm}  {\small G. Joyce and D. Montgomery, J. Plasma Phys. {\bf 10}, 107
(1973).}

\bibitem{kin}  {\small P.H. Chavanis, Phys. Rev. E {\bf 64}, 026309 (2001).}

\bibitem{ellis}  {\small R. Ellis, K. Haven and B. Turkington, Nonlinearity {\bf 15}, 239 (2002).}

\bibitem{bouchet}  {\small F. Bouchet, PhD thesis, Universit\'e Joseph Fourier (2001).}

\bibitem{biler}  {\small P. Biler and T. Nadzieja, Rep. Math. Phys. {\bf 52}, 205 (2003).}

\bibitem{rsmepp}  {\small R. Robert and J. Sommeria, Phys. Rev. Lett. {\bf 69}, 2776 (1992).}



\bibitem{dubrovnik}  {\small  Chavanis P.H., 2002, Statistical mechanics of violent relaxation in stellar systems. In: Proceedings of the Conference on Multiscale Problems in Science and Technology (Springer) [astro-ph/0212205]}

\bibitem{ks} {\small E. Keller and L.A. Segel, J. theor. Biol. {\bf 26}, 399 (1970). }

\bibitem{jager}  {\small W. J\"ager  and S. Luckhaus, Trans. Am. Math. Soc. {\bf 329}, 819 (1992).}





\end{thebibliography}
\end{document}